\documentstyle[twocolumn,pra,aps,epsfig]{revtex}

\begin{document}
\draft
\title{Controlling Decoherence in Bose-Einstein Condensation \\
by Light Scattering far off Resonance}
\author{C.P.Sun\cite{email,www}}
\address{Institute of Theoretical Physics,Academia Sinica,Beijing, 100080,China\\
(Feb. 19, 2000)\\
\medskip}
\author{\parbox{14.2cm}{\small \hspace*{3mm}It is shown
that, through a super-radiant Rayleigh scattering, a strong far
off-resonant pump laser applied to a Bose-Einstein
condensates(BEC) can induce a non-demolition coupling of the
many-mode quantized vacuum field to the BEC. This effective
interaction will force the total system of the BEC plus the light
field to evolve from a factorized initial state to an ideal
entangled state and thus result in the quantum decoherence in the
BEC. Since the effective coupling coefficients are mainly
determined by the Rabi frequency of the pump laser, the quantum
decoherence process can be controlled by adjusting the intensity
of the pump laser. To study the physical influence  of decoherence
on the BEC, we investigate how the coherent tunneling of BEC in a
well-separated tight double wall is suppressed by the
effectively-entangled vacuum modes. \vskip30pt PACS
numbers:03.65,03.75-b,42.50.-p,42.50.Ct,42.50.Dv}}
\maketitle

\flushbottom
\narrowtext
\vskip50pt

Recently, some important developments in Bose Einstein condensates (BEC) of
trapped atomic vapors \cite{AndEnsMat95,DavMewAnd95} have been achieved in
connection with light scattering of BEC, such as the Bragg scattering \cite
{deng}and Rayleigh scattering in various cases where the atoms interact only
with far off-resonant optical fields\cite{MIT}. The super-radiant Rayleigh
scattering by the BEC was observed in a recent experiment by Ketterle and
his co-workers at MIT. It was also demonstrated \cite{MM}\cite{Scat} that,
when the far off-resonant laser light is scattered into the vacuum modes of
the electro-magnetic field, the dominant two-photon interaction can
optically manipulate matter-wave coherence properties to generate a quai-CW
atomic laser and the so called atomic four-wave mixing from BEC \cite{deng}.
In this paper we will probe into the problem whether the far off-resonant
pump light scattering from a classical pumped mode to many-vacuum modes can
generate an ideal entanglement between atomic and optical fields and then
demolish the quantum coherence of the scattered condensate. Especially we
will consider how to use the pump laser to control the quantum decoherence
process of BEC\cite{decoh}.

Usually, an interaction between a quantum system and an environment can
cause two types of quantum irreversible effects: quantum dissipation
indicating the loss of the system energy, and quantum decoherence indicating
the leaking out of coherent information while the system energy is
conserved. In the study of irreversible process in the trapped BEC, progress
has been made by adopting particular forms of coupling between BEC atoms and
the environment,or in other words,by modeling the decoherence\cite{decoh}.
For BEC\ system, the natural origin of environment coupling is the
interaction between BEC\ atoms and the background electromagnetic field or
the thermal atomic cloud. It is noticed that this kind of interaction can
also change the energy of BEC and thus can also lead to a quantum
dissipation. So a pure decoherence without dissipation can not be modeled
directly based on the usual electromagnetic coupling. To detect the quantum
decoherence phenomenon clearly in a practical experiment, it is essential to
find a really-physical environment to produce quantum decoherence purely
without quantum dissipation. Fortunately, the above mentioned far
off-resonant light scattering on BEC can just induce such an environment
coupling to BEC\ system purely with decoherence. Since the effective
coupling coefficients are determined mainly by the Rabi frequency of the
pump laser, the quantum decoherence process can be controlled by adjusting
the intensity of the pump laser. To study the physical effect of decoherence
in the BEC, we consider how the coherent tunneling of BEC in a
well-separated tight double wall is suppressed by the effectively -entangled
vacuum modes.

Following Moore and Meystre et al \cite{MM}, we consider the Schreodinger
fields of two-level atoms coupled to a classical pump laser field far-off
resonant as well as to the quantized vacuum modes of the electromagnetic
field, via the electric-dipole interaction. For the large detuning between
the pump frequency and the atomic transition frequency, the excited state
field can be eliminated adiabatically and then all atoms can be described by
a scalar field $\hat{\Psi}_g({\bf r})$ of the ground state. The
corresponding effective Hamiltonian is given by\cite{MM} 
\begin{eqnarray}
\hat{{\cal H}} &=&\int d^3{\bf r}\hat{\Psi}_g^{\dag }({\bf r})H_A\hat{\Psi}%
_g({\bf r})+\sum_k\hbar \omega _ka_k^{\dagger }a_k+H_{AA}  \nonumber \\
&&+(\sum_k\hbar g_k\int d^3{\bf r}\hat{\Psi}_g^{\dag }({\bf r})\hat{\Psi}_g(%
{\bf r})a_ke^{i({\bf k-k}_0{\bf )}\cdot {\bf r}}+H.c),
\end{eqnarray}
where $H_{AA}=\hbar \xi \int d^3{\bf r[}\hat{\Psi}_g^{\dag }\hat{\Psi}%
_g^{\dag }\hat{\Psi}_g\hat{\Psi}_g]$ denotes two -body interatomic
interaction measured by the scattering length $a$ of $s-wave$ and $\xi =%
\frac{4\pi \hbar a}M$; $H_A$ =$-\frac{\hbar ^2}{2M}\nabla ^2+V({\bf r})$ is
the free Hamiltonian for single atom of mass $M$ in an effective trap $V(%
{\bf r})=V_g({\bf r})+\frac{\hbar R^2}{2\Delta }$; $a_k^{\dagger }$ and $a_k$
are the creation and annihilation operators for the vacuum electromagnetic
field with frequency $\omega _k=ck-\omega _0$ for $k=|{\bf k|}$, $c$ being
the speed of light. The effective trap $V({\bf r})$ was obtained from the
trap potential $V_g({\bf r})$ for the ground state with modification by the
coupling strength $R$ (Rabi frequency ) between the atoms under
consideration and the classical pump filed of frequency $\omega _0=ck_0$,
and $R(\propto \sqrt{I})$depends on the pump intensity $I$. With respect to
the atomic transition frequency $\omega _a$, the detuning $\Delta $ =$\omega
_0-\omega _a$ is very large. Particularly, it is noticed that the coupling
coefficient $g_k\propto \frac{|R|\sqrt{k}}{2|\Delta |}\propto \frac{\sqrt{Ik}%
}{2|\Delta |}$ \cite{MM}. So, in principle, one can control the quantum
dynamic processes of BEC resulting from the coupling to the vacuum, such as
the quantum decoherence.

When the BEC happens in a single trap $V_g({\bf r})$, the atomic field $\hat{%
\Psi}_g({\bf r})$ can be approximately quantized as\cite{M-BEC} $\hat{\Psi}%
_g({\bf r})\sim b_0\phi _0$ where $\phi _{0\text{ }}$is the wave function of
the ground state of energy $E_0=\hbar \Omega $ and $b_0$ is its annihilation
operator obeying $[b_0,b_0^{\dagger }]=1$. Then, the effective Hamiltonian
takes the form

\begin{eqnarray}
\hat{{\cal H}_e} &=&\hbar \Omega b_0^{\dagger }b_0+\hbar \kappa b_0^{\dagger
2}b_0^2+  \nonumber \\
&&\sum_{{\bf k}}\hbar \omega _{{\bf k}}a_{{\bf k}}^{\dagger }a_{{\bf k}%
}+b_0^{\dagger }b_0(\sum_{{\bf k}}\hbar \eta _{{\bf k}}a_{{\bf k}}+H.c),
\end{eqnarray}
where $\eta _{{\bf k}}=\hbar g_k\int d^3{\bf r}\phi _{0\text{ }}^{*}({\bf r}%
)\phi _{0\text{ }}({\bf r})e^{i({\bf k-k}_0{\bf )}\cdot {\bf r}}$ and $%
\kappa =\xi \int d^3{\bf r|}\phi _{0\text{ }}({\bf r})|^4$.The above
Hamiltonian describes a typical quantum decoherence process since the
effective interaction $H_I=b_0^{\dagger }b_0[\sum_{{\bf k}}\hbar \eta _{{\bf %
k}}a_{{\bf k}}+H.c]$ commutes with the free BEC Hamiltonian $H_0=\hbar
\Omega b_0^{\dagger }b_0+\hbar \kappa b_0^{\dagger 2}b_0^2$ , i.e., $%
[H_I,H_0]=0$\cite{scp}$.$ However, $H_I$ can still induce a virtual
transition, in which the atomic internal state remains unchanged. But it
should be noticed that since $[H_I,H_L]\neq 0$, this transition may result
in a back-action on the vacuum light field of the free Hamiltonian $%
H_L=\sum_{{\bf k}}\hbar \omega _{{\bf k}}a_{{\bf k}}^{\dagger }a_{{\bf k}}$ .

For example, when an atom in the Fock state $|n\rangle =\frac 1{\sqrt{n!}}%
b_0^{\dagger n}|n\rangle $ absorbs a pump photon and then emits a pump
photon, the vacuum light field experiences different recoil kicks described
by $H_n=n(\sum_{{\bf k}}\hbar \eta _{{\bf k}}a_{{\bf k}}+H.c)$ where $n$ is
the atom number. After time $t$ , this external forced term will drive the
vacuum light field to evolve into a product state $|v_n\rangle =\Pi _{{\bf k}%
}e^{i\gamma _{nk}(t)}|\alpha _{{\bf k}}^n\rangle $ of coherent states $%
|\alpha _{{\bf k}}^n(t)\rangle =\exp [\alpha _{{\bf k}}^n(t)a_{{\bf k}%
}^{\dagger }-H.c]|0\rangle $ where\cite{scp} 
\begin{eqnarray}
\alpha _{{\bf k}}^n(t) &=&\frac{ng_{{\bf k}}}{i\omega _{{\bf k}}}%
(e^{-i\omega _{{\bf k}}t}-1])  \nonumber \\
\gamma _{n{\bf k}} &=&\frac{n^2|g_{{\bf k}}|^2}{\omega _{{\bf k}}^2}[\omega
_{{\bf k}}t-\sin (\omega _{{\bf k}}t)]
\end{eqnarray}

Therefore, the total system formed by the BEC plus the vacuum field can
evolve from a factorized initial state $|\Psi (0)\rangle
=\sum_{n=0}c_n|n\rangle \otimes |0\rangle $ to an entangled state 
\begin{equation}
|\Psi (t)\rangle =\sum_{n=0}c_ne^{-i\epsilon (n)t}|n\rangle \otimes
|v_n\rangle
\end{equation}
where $\epsilon (n)=n(\Omega _0-\kappa )+\kappa n^2$ is the Hartree-Fock
energy of the BEC. The overlaps $O_{mn}=\langle v_m|v_n\rangle $, which are
called decoherence factors, can be calculated directly as follows 
\begin{eqnarray}
O_{mn} &=&\prod_{{\bf k}}\langle \alpha _{{\bf k}}^m|\alpha _{{\bf k}%
}^n\rangle  \nonumber \\
&=&\exp [-\sum_{{\bf k}}(m-n)^2\frac{g_{{\bf k}}^2}{\omega _{{\bf k}}^2}\sin
^2(\frac{\omega _{{\bf k}}t}2)]\times \\
&&\exp [\sum_{{\bf k}}i(m^2-n^2)\frac{g_{{\bf k}}^2}{\omega _{{\bf k}}^2}%
(\omega _{{\bf k}}t-\sin (\omega _{{\bf k}}t))]  \nonumber
\end{eqnarray}
It characterizes the extent of entanglement and has a factorized structure 
\cite{scp}with respect to the individual mode ${\bf k}$ of the
''environment''. All necessary information about the decoherence effects of
the vacuum environment on the BEC are contained in the decoherence factor.
The zero overlap means an ideal entanglement and thus leads to a complete
decoherence. In fact, if $\langle v_m|v_n\rangle =0$ for $m\neq n,$the
reduced density $\rho $ matrix of atoms have the vanishing off-diagonal
elements $\rho _{mn}=e^{i[\epsilon (m)-\epsilon (n)]t}\langle v_m|v_n\rangle
\rightarrow 0$.

Using the explicit expression of $g_{{\bf k}}$ given in ref.\cite{MM}, we
replace the sum $\sum_{{\bf k}}$ $(...)$ in Eq.(6) with an integral $\int $ $%
\mu (k)(...)k^2\sin \theta d\theta d\phi dk$ where$\theta $ and $\phi $ are
the polar variables of the wave vector ${\bf k}$. For the isotropic spectral
distribution (effective mode density) of the vacuum field, we can explicitly
calculate the norm of the decoherence factors:

\begin{eqnarray}
|O_{mn}(t)| &=&\exp [-(m-n)^2\frac{\pi R^2cd^2}{8(2\pi )^3\hbar \Delta ^2}%
\times  \nonumber \\
&&\int_0^\infty \frac{\sin ^2(\frac t2[ck-\omega _0])k^3}{(ck-\omega _0)^2}%
\mu (k)dk]\times
\end{eqnarray}
In the free space with $\mu (k)=1$, the integral in the above Eq.(6)
diverges to positive infinity and thus $O_{mn}(t)$ approaches zero. This
leads to the instantaneous quantum decoherence of the BEC. However, if we
put the system into a micro-cavity, the effective mode density $\mu (k)$
will change singularly\cite{QED}. In this case $\mu (k)$ may take certain
special forms so that the integral does not diverge to positive infinity.
For instance, in a special case where $\mu (k)=\frac \xi {k^3},$

\begin{equation}
|O_{mn}|=e^{-\lambda _{mn}[\pi t-2+2\omega _0^{-1}\cos (\omega _0t)+2%
\mathop{\rm Si}%
\left( \omega _0t\right) t]}
\end{equation}
where $\lambda _{mn}=\frac{\xi (m-n)^2R^2d^2}{256\pi ^2\hbar \Delta ^2}$ and 
$%
\mathop{\rm Si}%
(z)=\allowbreak \int_0^z$ $\frac{\sin x}xdx$ is a special function.
Figure.1(a-b) illustrates the decoherence processes for different $\lambda
_{mn}$ and $\omega _0.$What is of high interest is the fact that the damping
rate of quantum coherence described by $\rho _{mn}$ $\propto O_{mn}$ depends
on both the frequency $\omega _0$ of the pump light field and its coupling
strength $R$ to the BEC. Thus there arises a possibility for one to control
the happening of the quantum decoherence.

On the other hand, the leading damping term $\lambda _{mn}\pi $=$\frac{\xi
(m-n)^2R^2d^2}{256\pi \hbar \Delta ^2}$ , which plays a role in the
decoherence process for much longer time , does not depends on $\omega _0$,
so in a long time scale only different $R$ (or $\lambda _{mn}$) can notably
change the decaying behavior of quantum coherence. Figure.1a shows this
result qualitatively. It is clearly seen in Figure .1b that the curves $%
|O_{mn}(t)|$ almost remain unchanged for different $\omega _0$. Therefore,
one can artificially control the quantum dynamic process of decoherence for
the BEC by adjusting the physical parameter, namely, the intensity of the
classical pump laser.

As shown above quantum decoherence can only happen for the superposition of
the Fock states of different atomic numbers. If we understand BEC state as a
Fock state with definite atomic number $N$, no quantum decoherence happens
and the BEC is certainly robust. However, there exists a different view: an
assembly of the BEC atoms must be assigned a definite phase and so the BEC
is not a number state though it has an average atomic number $N$\cite{c-bec}%
. From this viewpoint a good pure state description for BEC is obtained by
using the coherent state,which can survive in an usual open environment much
longer than the number state does\cite{c-bec}. Since the BEC coherent state
is a superposition of atomic number states, the quantum decoherence can be
induced by the Rayleigh scattering far-off-resonance. By testing the
different effects of decoherence for the number state and the coherent
state, we can judge which of the two views concerning BEC reflects the
physical reality better.

To further analyses the influences of decoherence on BEC, we consider atomic
tunneling in the BEC formed in a symmetric double-well atomic trap\cite
{M-BEC}, whose potential $V({\bf r})$ has two well separated minima at $%
x=\pm \frac a2$. Assume the potential is such that the two lowest states $%
\phi _0({\bf r)}$ and $\phi _1({\bf r)}$ with even and odd parities are
closely spaced and well separated from higher levels. Their energy
difference $\hbar \delta \propto \frac 1{M\xi }e^{-2\xi a}$ where $\xi =%
\sqrt{2MV_0}$depends on the height of maxima of the double-well potential.
In this case with very large $\xi ,$ the interactions with the atoms in
higher levels do not significantly change the dynamics of atoms in the two
lowest states. Thus, a two-mode approximation is permitted for the many-body
description of the BEC\ system\cite{M-BEC}. In this way we can quantize the
atomic fields as 
\begin{equation}
\hat{\Psi}_g({\bf r})\sim b_0\phi _0+b_1\phi _1=b_r\phi _r+b_l\phi _l
\end{equation}
where we have introduced the annihilation operators $b_{r,l}$ =$\frac 1{%
\sqrt{2}}(b_0\pm $ $b_1{\bf )}$ for the right and the left local modes $\phi
_{r,l}({\bf r)=}\frac 1{\sqrt{2}}[\phi _0({\bf r)}$ $\pm \phi _1({\bf r)]}$.
If the position uncertainty in the states $\phi _l({\bf r)}$ and $\phi _r(%
{\bf r)}$ is much less than the separation of the minima of the global
potential $V({\bf r})$ , the local modes may be treated as orthogonal modes
and $[b_r,b_l^{\dagger }]\sim 0$. Then, neglecting the interatomic
interaction we can write down the effective Hamiltonian

\begin{eqnarray}
\hat{{\cal H}_e} &=&\hbar \Omega {\bf N}+{\bf \hbar \delta T+}\sum_{{\bf k}%
}\hbar \omega _{{\bf k}}a_{{\bf k}}^{\dagger }a_{{\bf k}}+  \nonumber \\
&&(\sum_{{\bf k}}\hbar (\mu _{{\bf k}}{\bf N+\zeta }_k{\bf T})a_{{\bf k}%
}+H.c),
\end{eqnarray}
for the double-well BEC scattered by a far-off resonance light. Here, ${\bf %
N=}b_l^{\dagger }b_l+b_r^{\dagger }b_r$ is the total atomic number and ${\bf %
T=}b_lb_r^{\dagger }+b_rb_l^{\dagger }$ is the tunnelling operator between
the two minima. The corresponding tunneling frequency is just the transition
frequency ${\bf \delta }$ of the two non-local motional modes $\phi _0$ and $%
\phi _1$. Here, the effective coupling constants $\mu _{{\bf k}}$ and ${\bf %
\zeta }_k$ are defined by $\mu _{{\bf k}}=\hbar g_k\int d^3{\bf r}\phi _{l%
\text{ }}^{*}({\bf r})\phi _{l\text{ }}({\bf r})e^{i({\bf k-k}_0{\bf )}\cdot 
{\bf r}}$ and ${\bf \zeta }_k=\hbar g_k\int d^3{\bf r}\phi _{l\text{ }}^{*}(%
{\bf r})\phi _{r\text{ }}({\bf r})e^{i({\bf k-k}_0{\bf )}\cdot {\bf r}}$ .

If the BEC system is initially prepared in the left local mode, we describe
it by a coherent state $|\alpha \rangle _l=\sum_{n=0}^\infty c_n(0)|n\rangle
_l$where $c_n=e^{-\frac 12|\alpha |^2}\frac{\alpha ^n}{\sqrt{n!}}$.
Interacting with the vacuum modes of the electromagnetic fields, the BEC
system plus the vacuum field will be driven from the factorized state $|\Psi
(0)\rangle =\sum_{n=0}c_n|n\rangle _l\otimes |0\rangle _r\otimes |0\rangle $
($|0\rangle _r$ denotes the vacuum of the right local mode in the double
well) into an entangled state 
\[
|\Psi (t)\rangle =\sum_{n=0}^\infty \sum_{m=0}^nc_ne^{-in(\Omega +\delta
)t}f_m^n(t)|n-m,m\rangle \otimes |v_m^n(t)\rangle 
\]
where 
\[
|m,k\rangle =\frac 1{\sqrt{m!k!}}b_1^{\dagger m}b_0^{\dagger k}|0\rangle 
\]
\begin{equation}
f_m^n(t)=\frac{(-1)^m}{(\sqrt{2})^n}\sqrt{\frac{n!}{(n-m)!m!}}e^{-2ik\delta
t}
\end{equation}
\[
|v_m^n(t)\rangle =\prod_ke^{-it[\omega _{{\bf k}}a_{{\bf k}}^{\dagger }a_{%
{\bf k}}-[\mu _{{\bf k}}n{\bf +\zeta }_k(n-2m)a_{{\bf k}}+H.c)]}|0\rangle 
\]

Based on the above exact solution, we now consider the atomic tunneling
between the two condensates in the presence of decoherence. The atomic
tunneling is usually characterized by the population difference between two
condensates 
\begin{eqnarray}
p(t) &=&\langle b_r^{\dagger }b_r-b_l^{\dagger }b_l\rangle =2%
\mathop{\rm Re}%
(\langle b_1^{\dagger }b_0\rangle ) \\
&=&2%
\mathop{\rm Re}%
\{\sum_{n=0}^\infty \sum_{m=0}^{n-1}|c_n|^2f_m^nf_{m+1}^{n*}\sqrt{(m+1)(n-m)}%
O_m\}  \nonumber
\end{eqnarray}
where $O_m=\langle v_{m+1}^n(t)|v_m^n(t)\rangle $is the decoherence factor
for the entangling vacuum fields. According to Eq.(9), we can roughly write $%
O_m\approx Je^{imS}$ \cite{scp}for the real time-dependent functions $J(t)$
and $R(t)$ and then calculate out the population difference in a compact form

\begin{eqnarray}
p(t) &=&%
\mathop{\rm Re}%
(-J\alpha ^2e^{\frac 12\left( 1-e^{iR}\right) \alpha ^2}e^{2i\delta t}) 
\nonumber \\
&=&-J\alpha ^2\cos (2\delta t)%
\mathop{\rm Re}%
e^{\frac 12\left( 1-e^{iS}\right) \alpha ^2}+ \\
&&-J\alpha ^2\sin (2\delta t)%
\mathop{\rm Re}%
(ie^{\frac 12\left( 1-e^{iS}\right) \alpha ^2})  \nonumber
\end{eqnarray}
where $J(t)=|O_m|$ is a damping factor similar to that in Eq.(6).

The above result answers the following natural question: what is the effect
of decoherence on the quantum coherent atomic tunneling ? It indicates that
the decoherence always tends to suppress the atomic tunneling current
between the two condensates since $J(t)$ is decaying with time $t.$ The
similar observation was even made by Kuang et.al most recently\cite{decoh},
but our result seems to be more clearly. Without quantum decoherence ( $%
O_m=1)$ the atomic tunneling manifests a simple harmonic oscillation, $%
p(t)=\alpha ^2\cos (2\delta t).$ Generally it is modified with a decaying
factor $J(t)$ and a phase shift $\theta $ defined by 
\begin{equation}
\tan \theta =\frac{%
\mathop{\rm Re}%
e^{\frac 12\left( 1-e^{iS}\right) \alpha ^2}}{%
\mathop{\rm Re}%
(ie^{\frac 12\left( 1-e^{iS}\right) \alpha ^2})}
\end{equation}
In Figure 2 we show this modification by plotting the time-dependent curves
of $p(t)$ for different $\theta $ and $J(t)$.

We have shown that a strong far off-resonant pump laser applied to a BEC can
produce quantum entanglement between the BEC and the many-mode quantized
vacuum field. It thus can result in quantum decoherence in the BEC. One of
the most interesting indications of this observation is the possibility to
control the dynamic processes of the BEC\ decoherence in the BEC by
engineering the coupling between the BEC and its environment. We notice the
effective coupling coefficients $g_{{\bf k}}$ or $\mu _{{\bf k}}$ and ${\bf %
\zeta }_k$ are determined by the Rabi frequency $R$ of the pump laser and
the pump frequency $\omega _0.$ Therefore, the quantum decoherence process
can be controlled by adjusting the intensity of the pump laser and the pump
frequency $\omega _0$. It can be seen from Eq.(6) that the norm of the
decoherence factor is more sensitive to the intensity of the pump laser than
to the pump frequency $\omega _0$. So the dominant element governing the
induced quantum decoherence of the BEC is the intensity of the pump laser
rather than the pump frequency $\omega _0$. The above discussion in this
paper is echoed by current experiments. For instance, a recent experiment by
Ketterle's group studied a condensate driven by a far-off resonant pump
laser. Its generalization maybe clearly demonstrate many aspects of the
present theoretical investigation for engineering the decoherence theory of
BEC. In fact, certain engineered environments have been implemented recently
to observe the decoherence of ion and cooled atom systems quantitatively\cite
{winland}.

There also exists another way to engineer the induced environment around the
BEC so that the quantum decoherence can be controlled effectively, that is,
to put the system in a an optical micro-cavity so that the spectral density $%
\mu (k)$ of the vacuum fields can be changed dramatically by the boundary of
cavity. In fact\cite{QED}, as a classic aspect of cavity quantum
electromagnetic theory, the suppression and enhancement of spontaneous
emission of atom within the cavity has been studied in many theoretical and
experimental investigations. Since the quantum dissipation caused by
spontaneous emission can be controlled by the cavity, it is natural to
expect that quantum decoherence,which is another quantum irreversible
process, can also be controlled . We will study,in the future research, the
quantum decoherence of the BEC\ system under the influence of cavity field
with various spectral density in details.

\acknowledgements
This work is supported in part by the National Foundation of Natural Science
of China.

\end{document}